# Solving Blind Inverse Problems: Adaptive Diffusion Models for Motion-corrected Sparse-view 4DCT

Antoine De Paepe[1], Alexandre Bousse[1], Clémentine Phung-Ngoc[1], and Dimitris Visvikis[1]

[1]Univ. Brest, LaTIM, INSERM, UMR 1101, 29238 Brest, France.

**Abstract** Four-dimensional computed tomography (4DCT) is essential for medical imaging applications like radiotherapy, which demand precise respiratory motion representation. Traditional methods for reconstructing 4DCT data suffer from artifacts and noise, especially in sparse-view, low-dose contexts. Motion-corrected (MC) reconstruction is a blind inverse problem that we propose to solve with a novel diffusion model (DM) framework that calibrates an adaptive unknown forward model for motion correction. Furthermore, we used a wavelet diffusion model (WDM) to address computational cost and memory usage. By leveraging the prior probability distribution function (PDF) from the DMs, we enhance the joint reconstruction and motion estimation (JRM) process, improving image quality and preserving resolution. Experiments on extended cardiac-torso (XCAT) phantom data demonstrate that our method outperforms existing techniques, yielding artifact-free, high-resolution reconstructions even under irregular breathing conditions. These results showcase the potential of combining DMs with motion correction to advance sparse-view 4DCT imaging.

## 1 Introduction

Four-dimensional computed tomography (4DCT) has become an indispensable tool in medical imaging, particularly for applications that require precise information about respiratory-induced motion, such as in radiotherapy planning. This imaging technique involves acquiring computed tomography (CT) scans at different couch positions throughout the respiratory cycle. In conventional reconstruction methods, these scans sorted and stacked according to surrogates signal to create a three-dimensional (3-D) image for each respiratory phase, typically 10 in total. However, irregular breathing can introduce artifacts into the reconstructed images of each respiratory phase. In addition, the reconstructed images suffer from noise amplification due to the low counts resulting from the gating process.

Various methods have been proposed to handle these artifacts. McClelland et al. [1] proposed a surrogate-driven motion model estimating a motion-free reference image and motion fields for each time point from unsorted CT scans. As the surrogate signals are not always available after acquisition, Huang et al. [2] proposed and extension of the previous works by considering surrogate signals as hyper-parameters to optimize.

Despite significant advancements in 4DCT motion artifact reduction, previous methods still face challenges when dealing with sparse-view data in low-dose contexts, as these algorithms typically operate in image space rather than in sinogram space.

In parallel, diffusion models (DMs) have emerged as a powerful tool for solving inverse problems [3], showing promising performance in medical imaging tasks such as image reconstruction [4]. Recent advancements have extended their application to blind inverse problems [5], where the forward operator is unknown, further showcasing their versatility.

In this paper, we explore the use of DMs within a blind inverse problem framework as a novel approach to motion-corrected (MC) sparse-view 4DCT. Our method achieves joint reconstruction and motion estimation (JRM) by employing an adaptive diffusion model (ADM) framework in which the forward operator integrates a surrogate-optimized motion model [2]. Furthermore, in order to address computational time and memory cost, we used the wavelet diffusion model (WDM) proposed by Friedrich et al. [6].

The rest of the paper is organized as follows. Section 2 introduces the forward problem in 4DCT followed by the corresponding blind inverse problem, and proposes a MC technique based on DMs. Section 3 shows our results on extended cardiac-torso (XCAT) phantom data. Section 4 discusses the limitations of our approach and proposes potential research directions. Finally, Section 5 concludes this work.

## 2 Materials and Methods

In the following, the 3-D attenuation image is represented by a vector $\boldsymbol{x} \in \mathbb{R}^m \triangleq \mathcal{X}$ with $m = n_\mathrm{p} \cdot n_z$ voxels, $n_\mathrm{p} = n_x \cdot n_y$ being the number of pixels per two-dimensional slices (e.g., $128^2$, $512^2$, etc.) and $n_z$ being the total number of slices. Furthermore, a deformation is represented by a 3-D deformation vector field (DVF) $\boldsymbol{\varphi} \in \mathbb{R}^{3 \times m}$, and we define $\mathcal{W}_{\boldsymbol{\varphi}} \colon \mathcal{X} \to \mathcal{X}$ as the corresponding image-to-image deformation operator.

### 2.1 Problem Formulation: Motion Model from Unsorted CT Scans

A 4DCT scan involves counting photons arriving at each detector across different view angles for a limited number of slices of the whole volume. The process is repeated $n_\mathrm{t}$ times at varying couch positions at different instants $\tau_k$, $k = 1, \ldots, n_\mathrm{t}$. During the acquisition, the 3-D image $\boldsymbol{x}$ is affected by respiratory motion, which is represented by a collection of DVFs $\{\boldsymbol{\varphi}_k\}_{k=1}^{n_\mathrm{t}} \in \mathcal{G}^{n_\mathrm{t}}$, and $\mathcal{W}_{\boldsymbol{\varphi}_k}(\boldsymbol{x})$ represents the deformed 3-D image at time index $k$. In order to reduce the number of parameters, we proceed in a similar fashion to McClelland et al. [1] by assuming that the DVFs $\{\boldsymbol{\varphi}_k\}_{k=1}^{n_\mathrm{t}}$ are defined as a generic DVF $\boldsymbol{\phi} \in \mathcal{G}$ and a surrogate signal $\boldsymbol{s} = [s_1, \ldots, s_{n_\mathrm{t}}] \in \mathbb{R}^{n_\mathrm{t}} \triangleq \mathcal{S}$ as

$$\boldsymbol{\varphi}_k = s_k \cdot \boldsymbol{\phi} \quad \forall k. \qquad (1)$$

The entire 4DCT measurement is denoted $\boldsymbol{y} = \{\boldsymbol{y}_k\}_{k=1}^{n_\mathrm{t}} \in \mathcal{Y}^{n_\mathrm{t}}$, $\mathcal{Y} \triangleq \mathbb{R}^n$, where for all $k$ the vector $\boldsymbol{y}_k = [y_{1,k}, \ldots, y_{n,k}]^\top \in \mathcal{Y}$ represents the measurement at time $\tau_k$ and $n = n_\mathrm{s} \cdot n_\theta \cdot n_\mathrm{d}$ with $n_\mathrm{s}$, $n_\theta$ and $n_\mathrm{d}$ denoting respectively the number of slices of each sub-measurement, the number of view angles and the number of detectors. At each time $\tau_k$, the system acquires data corresponding to $n_\mathrm{s}$ slices of the warped volume $\mathcal{W}_{\boldsymbol{\varphi}_k}(\boldsymbol{x})$, and we define the associated slice extractor as $\mathcal{T}_k \colon \mathcal{X} \to \mathcal{X}' \triangleq \mathbb{R}^{m'}$.

The photon counting process is modeled with a Poisson random probability distribution function (PDF), i.e., for all

$i = 1, \ldots, n$ and for all $k = 1, \ldots, n_t$,

$$y_{i,k} \mid \boldsymbol{x}, \boldsymbol{\phi}, \boldsymbol{s} \sim \text{Poisson}(\bar{y}_{i,k}(\boldsymbol{x}, \boldsymbol{\phi}, \boldsymbol{s})) \quad (2)$$

where the conditional expectation $\bar{y}_{i,k}(\boldsymbol{x}, \boldsymbol{\phi}, \boldsymbol{s}) \triangleq \mathbb{E}[y_{i,k}|\boldsymbol{x}, \boldsymbol{\phi}, \boldsymbol{s}]$ is given by the Beer-Lambert law, i.e.,

$$\bar{y}_{i,k}(\boldsymbol{x}, \boldsymbol{\phi}, \boldsymbol{s}) = I \cdot e^{-[\mathcal{R} \circ \mathcal{T}_k \circ \mathcal{W}_{\boldsymbol{\varphi}_k}(\boldsymbol{x})]_i}, \quad \boldsymbol{\varphi}_k = s_k \cdot \boldsymbol{\phi}, \quad (3)$$

$\mathcal{R}\colon \mathcal{X}' \to \mathcal{Y}$ being the slice-by-slice fan-beam line integral operator and $I$ being the photon emission intensity.

In absence of prior on $(\boldsymbol{\phi}, \boldsymbol{s})$, MC reconstruction of the image $\boldsymbol{x}$ from the measurement $\boldsymbol{y}$ can be achieved by performing JRM though a maximum *a posteriori* optimization problem

$$\max_{\boldsymbol{x} \in \mathcal{X}, \boldsymbol{\phi} \in \mathcal{G}, \boldsymbol{s} \in \mathcal{S}} p(\boldsymbol{y} \mid \boldsymbol{x}, \boldsymbol{\phi}, \boldsymbol{s}) \cdot p(\boldsymbol{x}) \quad (4)$$

where the conditional PDF $p(\boldsymbol{y}|\boldsymbol{x}, \boldsymbol{\phi}, \boldsymbol{s})$ is given by (2) and (3) and $p(\boldsymbol{x})$ is the prior distribution on $\boldsymbol{x}$. An approximate solution of (4) is usually obtained using a penalized weighted least squares formulation. Defining the MC system matrix as $\mathcal{A}_{\boldsymbol{\phi},\boldsymbol{s}} \triangleq \left[ \mathcal{A}_{\boldsymbol{\phi},\boldsymbol{s}}^{1\top}, \cdots, \mathcal{A}_{\boldsymbol{\phi},\boldsymbol{s}}^{n_t\top} \right]^\top \colon \mathcal{X} \to \mathcal{Y}^{n_t}$ where $\mathcal{A}_{\boldsymbol{\phi},\boldsymbol{s}}^k = \mathcal{R} \circ \mathcal{T}_k \circ \mathcal{W}_{s_k \cdot \boldsymbol{\phi}} \colon \mathcal{X} \to \mathcal{Y}$, the negative log-posterior is approximated as (see Elbakri et al. [7])

$$-\log p(\boldsymbol{y} \mid \boldsymbol{x}, \boldsymbol{\phi}, \boldsymbol{s}) \approx \frac{1}{2} \|\mathcal{A}_{\boldsymbol{\phi},\boldsymbol{s}}(\boldsymbol{x}) - \boldsymbol{b}\|_{\boldsymbol{W}}^2 \quad (5)$$

where $\boldsymbol{b} = \{\boldsymbol{b}_k\}_{k=1}^{n_t} \in \mathcal{Y}^{n_t}$, $\boldsymbol{b}_k = [b_{1,k}, \ldots, b_{n,k}]^\top$ with $b_{i,k} \triangleq \log I/y_{i,k}$, and $\boldsymbol{W} \in \mathbb{R}_+^{n \cdot n_t \times n \cdot n_t}$ is a diagonal matrix of statistical weights. An alternative to (4) is therefore the standard JRM approach

$$\min_{\boldsymbol{x} \in \mathcal{X}, \boldsymbol{\phi} \in \mathcal{G}, \boldsymbol{s} \in \mathcal{S}} \frac{1}{2} \|\mathcal{A}_{\boldsymbol{\phi},\boldsymbol{s}}(\boldsymbol{x}) - \boldsymbol{b}\|_{\boldsymbol{W}}^2 + \gamma R(\boldsymbol{x}) \quad (6)$$

where $R\colon \mathcal{X} \to \mathbb{R}$ is a convex regularizer that replaces the unknown prior $-\log p(\boldsymbol{x})$ and $\gamma > 0$ is a weight.

## 2.2 Joint Reconstruction and Motion Estimation with Diffusion Models

### 2.2.1 Background on Diffusion Models

In absence of a tractable prior PDF $p(\boldsymbol{x})$, $\boldsymbol{x}$ can be sampled through a model trained through diffusion. A commonly adopted approach is the denoising diffusion probabilistic model [8], which samples $\boldsymbol{x}_t$ given $\boldsymbol{x}_{t-1}$, $t = 1, \ldots, T$, starting from an initial image $\boldsymbol{x}_0$ sampled from the training dataset with PDF $p^{\text{data}}$,

$$\boldsymbol{x}_t \mid \boldsymbol{x}_{t-1} \sim \mathcal{N}(\sqrt{\alpha_t}\boldsymbol{x}_{t-1}, (1-\alpha_t)\boldsymbol{I}_\mathcal{X}) \quad (7)$$

where $\boldsymbol{I}_\mathcal{X}$ is the identity matrix on $\mathcal{X}$ and $\alpha_t$ is a scaling factor such that $\boldsymbol{x}_T \sim \mathcal{N}(\boldsymbol{0}_\mathcal{X}, \boldsymbol{I}_\mathcal{X})$. One prominent sampling algorithm, denoising diffusion implicit model (DDIM) [9], approximates the reverse process and enables sampling an image from a generalized version of $p^{\text{data}}(\boldsymbol{x})$ that approximates the theoretical prior $p(\boldsymbol{x})$. It adopts the update rules

$$\begin{aligned}\boldsymbol{x}_{t-1} &= \sqrt{\bar{\alpha}_{t-1}}\hat{\boldsymbol{x}}_{0|t} \\ &+ \sqrt{1 - \bar{\alpha}_{t-1} - \sigma_t^2} \cdot \frac{\boldsymbol{x}_t - \sqrt{\bar{\alpha}_t}\hat{\boldsymbol{x}}_{0|t}}{\sqrt{1-\bar{\alpha}_t}} + \sigma_t^2 \boldsymbol{\epsilon}_t \\ \boldsymbol{\epsilon}_t &\sim \mathcal{N}(\boldsymbol{0}_\mathcal{Z}, \boldsymbol{I}_\mathcal{Z}),\end{aligned} \quad (8)$$

where $\bar{\alpha}_k = \prod_{s=1}^k \alpha_s$ and $\hat{\boldsymbol{x}}_{0|t} \triangleq \mathbb{E}[\boldsymbol{x}_0|\boldsymbol{x}_t]$ is given by Tweedie's formula,

$$\hat{\boldsymbol{x}}_{0|t} = \frac{1}{\sqrt{\bar{\alpha}_t}}(\boldsymbol{x}_t + (1-\bar{\alpha}_t)\nabla \log p_t(\boldsymbol{x}_t)), \quad (9)$$

$p_t$ being the PDF of $\boldsymbol{x}_t$. As the score $\nabla \log p_t(\boldsymbol{x}_t)$ is untractable, $\hat{\boldsymbol{x}}_{0|t}$ is approximated through a neural network (NN) $\boldsymbol{x}_{\boldsymbol{\theta}}\colon \mathcal{X} \times [0,T] \to \mathcal{X}$ with parameter $\boldsymbol{\theta} \in \Theta$ trained to recover $\boldsymbol{x}_0$ from $\boldsymbol{x}_t$ as

$$\min_{\boldsymbol{\theta} \in \Theta} \mathbb{E}_{t,\boldsymbol{x}_0,\boldsymbol{x}_t}\left[\|\boldsymbol{x}_{\boldsymbol{\theta}}(\boldsymbol{x}_t, t) - \boldsymbol{x}_0\|_2^2\right], \quad (10)$$

where $t \sim \mathcal{U}[0,T]$, $\boldsymbol{x}_0 \sim p^{\text{data}}$, and $\boldsymbol{x}_t \sim \mathcal{N}(\sqrt{\bar{\alpha}_t}\boldsymbol{x}_0, (1-\bar{\alpha}_t)\boldsymbol{I}_\mathcal{X})$.

### 2.2.2 Diffusion Models in Wavelet Transform Domain

Applying DMs to 3-D medical imaging is challenging due to high computational cost and memory usage. Latent diffusion models take into account these challenges by operating in a "compressed" space. Recent works proposed to perform the diffusion in a the wavelet domain [6], significantly reducing memory usage during training and inference while achieving state-of-the-art performance. A discrete wavelet transform (DWT) $\mathcal{V}\colon \mathcal{X} \to \mathcal{V} \triangleq \mathbb{R}^{8 \times \frac{n_x}{2} \times \frac{n_y}{2} \times \frac{n_z}{2}}$ is utilized to decompose a 3-D image $\boldsymbol{x}$ into a 8-channel wavelet coefficients 3-D image $\boldsymbol{v}$ with half the spatial dimension of $\boldsymbol{x}$. Following this, a DM framework is employed in the latent space $\mathcal{V}$ to sample $\boldsymbol{v}$ from a sequence of variables $\boldsymbol{v}_t$ by means of an NN $\boldsymbol{v}_{\boldsymbol{\theta}}\colon \mathcal{V} \times [0,T] \to \mathcal{V}$ trained to recover $\boldsymbol{v}_0$ from $\boldsymbol{v}_t$ in a similar fashion as for $\boldsymbol{x}_{\boldsymbol{\theta}}$ in (10).

### 2.2.3 Diffusion Posterior Sampling for Blind Inverse Problems

DMs can be used for image reconstruction via diffusion posterior sampling (DPS) using the conditional score

$$\nabla_{\boldsymbol{v}_t} \log p(\boldsymbol{v}_t|\boldsymbol{y}) = \nabla_{\boldsymbol{v}_t} \log p_t(\boldsymbol{v}_t) + \nabla_{\boldsymbol{v}_t} \log p(\boldsymbol{y}|\boldsymbol{v}_t). \quad (11)$$

Using the conditional score (11) for the image update (8) in the wavelet domain, defines a DPS approach to sample $\boldsymbol{v}$ given $\boldsymbol{y}$. However this approach requires a known forward model $\mathcal{A}_{\boldsymbol{\phi},\boldsymbol{s}}$. We therefore propose an ADM which calibrates the forward model by estimating $(\boldsymbol{\phi}, \boldsymbol{s})$ through minimization of the approximated negative log-likelihood (5) as proposed by Bai et al. [10] for blind inverse problems, with an update of the form

$$(\hat{\boldsymbol{\phi}}, \hat{\boldsymbol{s}}) \leftarrow \arg\min_{\boldsymbol{\phi} \in \mathcal{G}, \boldsymbol{s} \in \mathcal{S}} \frac{1}{2} \|\mathcal{A}_{\boldsymbol{\phi},\boldsymbol{s}}(\hat{\boldsymbol{x}}_{0|t}) - \boldsymbol{b}\|_{\boldsymbol{W}}^2 \quad (12)$$

where $\hat{\boldsymbol{x}}_{0|t} \triangleq \mathcal{V}^{-1}(\hat{\boldsymbol{v}}_{0|t})$, which we solve used an RMSprop algorithm with 5 iterations. Furthermore, while adopting the manifold preserving guided diffusion shortcut approach [11] allows to derive the next estimate $\boldsymbol{v}_{t-1}$ in the compressed space, as the DWT is bijective, we propose to enforce data consistency in image domain $\mathcal{X}$ using the posterior update rule as Zhu et al. [12]:

$$\tilde{\boldsymbol{x}}_{0|t} \leftarrow \arg\min_{\boldsymbol{x} \in \mathcal{X}} \frac{1}{2}\|\mathcal{A}_{\hat{\boldsymbol{\phi}},\hat{\boldsymbol{s}}}(\boldsymbol{x}) - \boldsymbol{b}\|_{\boldsymbol{W}}^2 + \frac{\zeta_t}{2}\|\boldsymbol{x} - \hat{\boldsymbol{x}}_{0|t}\|_2^2 \quad (13)$$

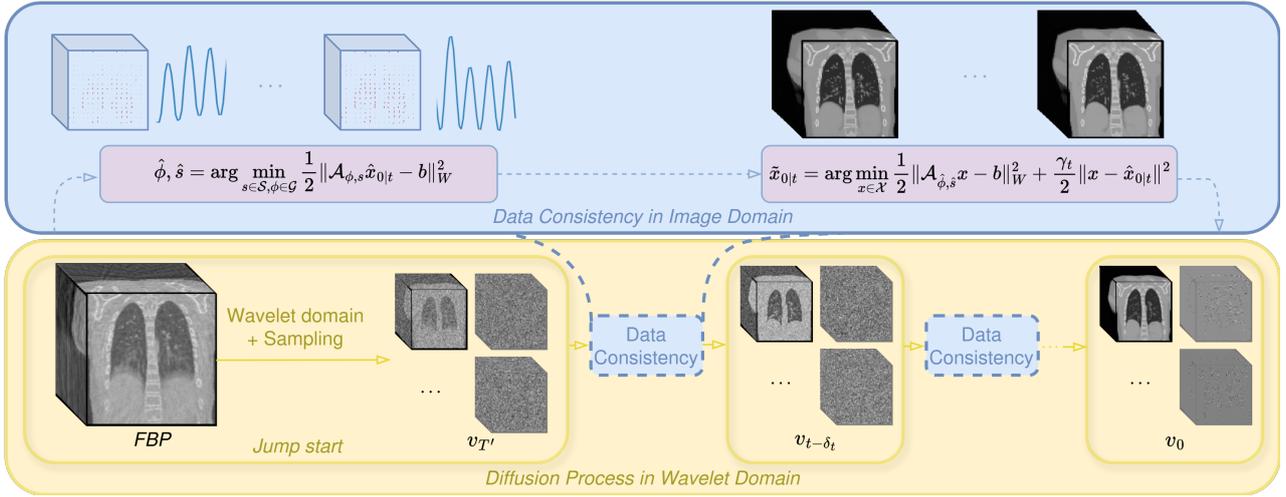

**Figure 1:** Summary of JRM-ADM.

which eliminates the need to backpropagate through $v_\theta$ and is solved with 5 RMSprop iterations. The new clean estimate $\tilde{v}_{0|t} \triangleq \mathcal{V}(\tilde{x}_{0|t})$ is then used to sample $v_{t-1}$ using (8) in the wavelet domain.

The overall method, which we name JRM-ADM, is summarized in Figure 1 and Algorithm 1.

To enhance stability, we used the jumpstart strategy proposed by Jiang et al. [13] using an initial image $x^{js}$ reconstructed slice by slice from gated data at end-inhale phase using filtered backprojection (FBP), which is then used to produce an initial wavelet coefficient image $v^{js} = \mathcal{V}(x^{js})$, thus allowing to start the sampling process from $T' < T$. In addition, we implemented the DDIM approach with $\sigma_t = 0$ and with a time step $\delta_t > 1$, as proposed in Song et al. [9]. Finally, we parametrize $\phi$ with B-splines and we used a standard sinusoidal signal to initialize $s$.

**Algorithm 1** Pseudo code of JRM-ADM.

**Require:** $T'$, $v^{js}$, $y$, $\{\zeta_t\}_{t=1}^{T'}$, $\{\delta_t\}_{t=1}^{T'}$, $\{\alpha_t\}_{t=t}^{T'}$
1: $t \leftarrow T'$
2: $\epsilon \sim \mathcal{N}(\mathbf{0}_\mathcal{X}, \mathbf{I}_\mathcal{X})$
3: $v_t \leftarrow \sqrt{\bar{\alpha}_t} v^{js} + \sqrt{1-\bar{\alpha}_t} \epsilon$
4: **while** $t > 0$ **do**
5: $\quad \hat{v}_{0|t} \leftarrow v_\theta(v_t, t)$
6: $\quad \hat{x}_{0|t} \leftarrow \mathcal{V}^{-1}(\hat{v}_{0|t})$
7: $\quad (\hat{\phi}, \hat{s}) \leftarrow \arg\min_{\phi \in \mathcal{G}, s \in \mathcal{S}} \frac{1}{2} \|\mathcal{A}_{\phi,s}(\hat{x}_{0|t}) - b\|_W^2$
8: $\quad \tilde{x}_{0|t} \leftarrow \arg\min_{x \in \mathcal{X}} \frac{1}{2} \|\mathcal{A}_{\hat{\phi},\hat{s}}(x) - b\|_W^2 + \frac{\zeta_t}{2}\|x - \hat{x}_{0|t}\|_2^2$
9: $\quad \tilde{v}_{0|t} \leftarrow \mathcal{V}(\tilde{x}_{0|t})$
10: $\quad v_{t-\delta_t} \leftarrow \sqrt{\bar{\alpha}_{t-\delta_t}} \tilde{v}_{0|t} + \sqrt{1-\bar{\alpha}_{t-\delta_t}} \cdot \frac{v_t - \sqrt{\bar{\alpha}_t}\tilde{v}_{0|t}}{\sqrt{1-\bar{\alpha}_t}}$
11: $\quad t = t - \delta_t$
12: **end while**
13: **return** $\hat{x}_0, \hat{\phi}, \hat{s}$

## 3 Experiments

### 3.1 Training, 4DCT Data Simulation and Evaluation

All the reconstruction methods and simulations were implemented in Pytorch, and we used TorchRadon [14] for the CT fan-beam projector.

Approximately 200 attenuation phantoms with varying morphologies were generated using the XCAT software [15]. For each phantom, about 20 respiratory phases were generated to diversify the dataset. Each phantom consists of 3-D $128 \times 128 \times 128$ volumes with a 2.6-mm isotropic voxel size.

We trained the NN $v_\theta$ using the Adam optimizer with approximately 150 epochs. The training was performed on standardized volumes, and the standardization was taken into account in the forward model.

To evaluate our method, we generated five four-dimensional phantom with morphologies that differ from the training dataset, each of which consisting of a collection $\{x_k\}_{k=1}^{n_t}$ with $n_t = 170$ and comprising 17 respiratory cycles, featuring a mix of regular and irregular cycles. For each time step, the raw data $y_k$ was generated following (2) by replacing $\mathcal{W}_{\varphi_k} x$ by the true volume $x_k$.

The simulated 4DCT system acquires data corresponding to $n_s = 8$ slices at each time index $k$, with $n_d = 192$ detectors and $n_\theta = 52$ angles of view. The overall setting results in a pitch of 0.1, and we used a source intensity value of $I = 10^5$.

The evaluation is conducted on the end-inhale phase, which, on average, is the most affected by irregular breathing. We used the peak signal-to-noise ratio and structural similarity index measure as figures of merit, which were computed using a ground truth (GT) image $x_k$ corresponding to a regular end-inhale phase.

The reconstruction was jumpstarted from $T' = 300$ and DDIM was implemented with a time step $\delta_t = 10$.

For comparison, we implemented gated-FBP and gated-DPS, i.e., using a subset of $\{1, \ldots, n_t\}$ corresponding to the end-inhale phase. We also implemented standard JRM, i.e., by solving (6)—where we used a smoothed total variation (TV) penalty for $R$—with a limited-memory Broyden-Fletcher-Goldfarb-Shanno algorithm; this method is referred to as JRM-TV.

### 3.2 Results

GT and reconstructed images are shown in Figure 2. Gated-FBP exhibits pronounced noise and streak artifacts, whereas gated-DPS effectively mitigates these issues through the use

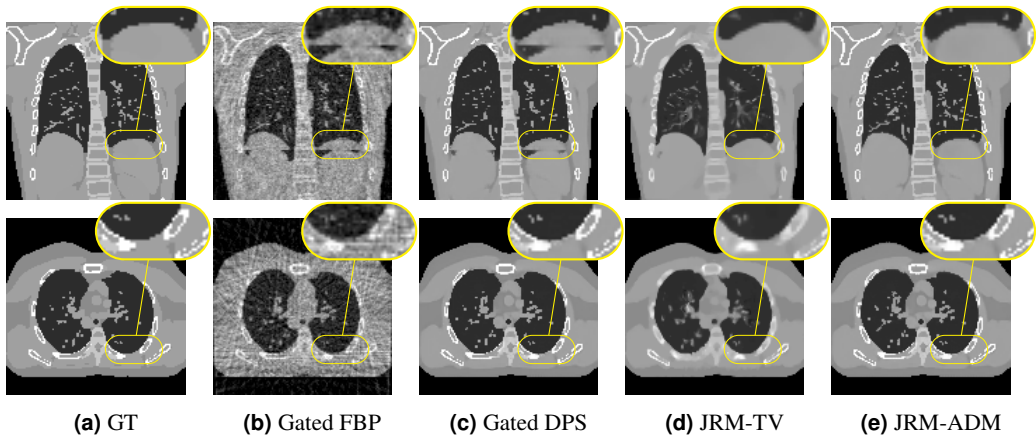

**(a)** GT   **(b)** Gated FBP   **(c)** Gated DPS   **(d)** JRM-TV   **(e)** JRM-ADM

**Figure 2:** GT and end-inhale phase reconstructions.

of the diffusion prior. However, they both suffer from motion artifacts around the diaphragm due to irregular breathing motion. JRM-TV produces noise-free and streak-free images, thanks to the TV regularization. Furthermore, JRM-TV avoids motion artifacts as it does not rely on gating. Nonetheless, it is limited by poor resolution, a consequence of the TV regularization. On the other hand, JRM-ADM produces noise- and artifact-free images while preserving the resolution.

This results are confirmed with the metrics (averaged over the five datasets) displayed in Table 1, showing that JRM-ADM outperforms all other methods.

|        | Gated FBP       | Gated DPS       | JRM-TV          | JRM-ADM          |
|--------|-----------------|-----------------|-----------------|------------------|
| PSNR ↑ | $20.59 \pm 0.27$ | $24.09 \pm 0.47$ | $25.04 \pm 0.49$ | **$27.05 \pm 0.37$** |
| SSIM ↑ | $0.37 \pm 0.01$  | $0.90 \pm 0.01$  | $0.89 \pm 0.01$  | **$0.94 \pm 0.01$**  |

**Table 1:** Quantitative evaluation (PSNR, SSIM) of four different reconstruction methods on the end-inhale phase for the five datasets.

## 4 Discussion

This study demonstrates the potential of combining motion correction and DMs in sparse-view 4DCT. However, several limitations remain to be addressed.

Firstly, computational time and memory usage are significant challenges. A possible solution is the patch-based approach proposed by Hu et al. [16]. Alternatively, our framework could be decomposed into a two-step process: first, by reconstructing each slab using DPS, and then by separately estimating the movement. However, this approach does not fully leverage the statistical noise model, which is crucial in the low-dose context of photon-counting computed tomography.

Secondly, our models were trained and evaluated on XCAT phantoms due to the limited availability of 4DCT datasets. Although these models demonstrate generalizability to unseen data [17], we are actively working to adapt them for use with real CT volumes. This involves developing a generative model to synthesize 4DCT images from static 3-D images, inspired by the recent work of Cao et al. [18].

## 5 Conclusion

This study introduces a novel framework that combines motion correction and DMs to address challenges in sparse-view 4DCT reconstruction. Our results highlight significant improvements in image quality compared to existing methods, with enhanced noise suppression, artifact reduction, and resolution preservation. Despite these advancements, challenges like high computational costs and limited training datasets remain. Future work will focus on optimizing the computational efficiency and expanding the model's applicability to real-world 4DCT data. This approach paves the way for more robust and accurate imaging techniques in clinical and research settings.

## Acknowledgement

This work was supported by CPER 2021–2027 IMAGIIS (INNOV-XS).